\documentstyle[12pt,epsf]{article}
\def\g{\gamma}
\def\m{\mu^2}
\def\G{\Gamma}
\def\gp{\g\pi^{\pm}}
\def\a{(\alpha-\beta)}
\def\b{(\alpha+\beta)}
\def\ab{\a_{\pi^0}}
\def\ac{\a_{\pi^{\pm}}}
\def\ap{\b_{\pi^{\pm}}}
\def\as{\b_{\pi^0}}
\def\gg{\g \g\to \pi^0 \pi^0}
\def\s{\sigma}
\def\sig{\frac{d\s_{\g\pi}}{d\Omega}}
\def\tg{\Theta^{cm}_{\g}}
\def\z{\cos\tg}
\def\mpp{M_{++}}
\def\mm{M_{+-}}
\def\sp{s'}
\def\tp{t'}
\def\be{\begin{equation}}
\def\ee{\end{equation}}
\def\beq{\begin{eqnarray}}
\def\eeq{\end{eqnarray}}
\begin{document}
\begin{center}
{\Large \bf Compton scattering on the charged pion \\
and the process $\gg$}\footnote{This work was supported by
the Deutsche Forschungsgemeinschaft (SFB 201)}
\vspace{0.5cm}

{\large \bf L.V. Fil'kov, V.L. Kashevarov}
\vspace{0.5cm}

{\it Lebedev Physical Institute, Leninsky Prospect 53,
Moscow 117924, Russia.\\ 
E-mail: filkov@x4u.lpi.ruhep.ru}
\end{center}
\vspace{1cm}

\begin{center}
Abstract
\end{center}
\vspace{0.2cm}

The Compton scattering on a charged pion  and the process $\gg$
are studied using the dispersion relations. Unknown parameters of
the $S$-wave $\pi\pi$ interaction and a sum and a difference of the
$\pi^0$ meson polarizabilities are found from a fit to the experimental
data for the $\gg$ process. The found parameters of the $\pi\pi$
interaction are used for the calculation of the cross section of
the elastic $\gp$ scattering. The analysis of the obtained results
shows that the experimental data for the elastic $\gp$ scattering
in the energy region up to 1 GeV together with the data for the $\gg$
process could be used both for a determination of the pion
polarizability values and for study
of the  $S$-wave $\pi\pi$ interaction.
\vspace{0.3cm}


\vspace{1cm}

\section{Introduction}

At present the experiments at Mainz \cite{ahr}, CEBAF \cite{nor}
and CERN \cite{moin} are planed to determine
the cross section of the elastic $\gp$ scattering in the energy region
$\sqrt{s}\le 420$ MeV (where $s$ is the square of the total energy in
$\g\pi$ c.m.s.) and extract from it
the pion polarizabilities. In the work \cite{dreck} it has been shown
that the corrections to the low energy expression for the cross section of
the Compton scattering on $\pi^{\pm}$ meson in the energy region under 
consideration
can be big. This could influence essentially the values of the
polarizability obtained from these experiments using the low energy
model. Therefore, it is necessary to find a correct theoretical
expression for the cross section allowed to extract the pion
polarizabilities with high precision in a wide enough energy region.
Moreover, it is of interest to
investigate what an information could be obtained from the analysis of
experimental data for this process at higher energy.

The aim of the present work is an investigation of the elastic
$\g\pi^{\pm}$ scattering in the energy region up to $\sqrt{s}\simeq 1$GeV
using the available experimental information about the process $\gg$.
For this purpose the dispersion relations (DRs) with the subtraction are
constructed for the $\g\pi$ scattering amplitudes. Unknown parameters
of the $S$-wave $\pi\pi$ interaction, represented as a broad
Breit-Wigner resonance, are determined from fits to the experimental data
\cite{mars} for the $\gg$ process for which the same DRs are used.

An investigation of the process $\g\g\to\pi\pi$ at low energies was
carried out in a number of works in the framework of the chiral
perturbation theory (ChPT). In the case of the charged pions, the
one-loop ChPT \cite{bij} describes quite well the experimental data
\cite{boy} at low energy whereas such an approximation is in disagreement
with the data for the $\gg$ process. The agreement with these data was
essentially improved by the two-loop calculation \cite{bell}.

In the work \cite{bur} it was shown that the two-loop contributions
induce only small changes in the cross section of the process
$\g\g\to\pi^+\pi^-$. The author of this work studied also the Compton
scattering on a charged pion and showed that the total cross section
of this process in the energy region up to $\sqrt{s}=350$ MeV is
dominated by the Born term and one- and two-loop corrections are
negligible.

The chiral perturbation approach is valid at low energies. To extend
the energy region, the authors of the work \cite{oset1}
gave a description of different channels of
$\g\g\to M\bar M$ reaction (where $M$ were $\pi$, $K$, $\eta$ mesons) using
for the final state interaction of mesons the results obtained in the
model \cite{oset2} which combines coupled channel Lippmann-Schwinger
equations with meson-meson potentials provided by the lowest order chiral
Lagrangian. Their results are in good agreement with the experimental
data for the $\g\g\to\pi^+\pi^-$ process up to $\sqrt{t}=1.4$ GeV.
However, in the case of the process $\gg$ their results lie significantly
higher than the experimental data \cite{mars} for
0.4GeV$<\sqrt{t}<$0.6GeV.

In work \cite{lee} the reaction $\g\g\to M\bar M$ was studied by using
the master formula approach to QCD with three flavors. To go out
beyond threshold region the authors used resonance saturation methods.

An analysis of the process $\g\g\to\pi\pi$ in the framework of DRs was
performed early in the works \cite{morg,kal,donog}. These DRs were
constructed for the $S$-wave amplitude and a correction for higher waves
was realized by taking into account poles of the vector mesons.

We construct the DRs at fixed $t$ with the subtraction for the invariant
amplitudes and
determine the subtraction functions through the DRs with the subtraction in
cross channels and the difference and the sum of the pion polarizabilities.
Such DRs allow to describe both the Compton scattering on the pion
and the process $\g\g\to\pi\pi$ simultaneously and to estimate expected
contributions of the pion polarizabilities in different  kinematical
regions of the process $\g+\pi^{\pm}\to\g+\pi^{\pm}$. The subtractions in
the DRs provide a good enough convergence of underintegral
expressions of these DRs and so increase the reliability of the
calculations.

The content of this work is as follows.
In Sec.2 the DRs for the amplitudes of $\g\pi$ scattering are constructed.
Determination of the parameters of the $S$-wave $\pi\pi$ interaction and
the $\pi^0$ meson polarizabilities are in Sec.3. In Sec.4 the
analysis of the Compton scattering on the charged pion is discussed and
one suggests a method of extraction of the pion polarizability from
the Compton scattering data in the energy region up to 1 GeV. The conclusions
are given in Sec.5.

\section{Dispersion relations for the amplitudes of $\gp$ scattering}

The Compton scattering on the pion is described by the following invariant
variables
\be
s=(p_1+k_1)^2, \quad u=(p_1-k_2)^2, \quad t=(k_1-k_2)^2
\ee
where $p_1(p_2)$ and $k_1(k_2)$ are the pion and photon initial (final)
4-momenta. The low energy expression for the cross section for Compton
scattering on charged pion \cite{dreck,filk1} is
\beq
&&\sig= \left(\sig\right)_B-\left(\frac{e^2}{4\pi}\right)
\frac{\mu^3(s-\m)^2}{4s^2[(s+\m)+(s-\m)z]} \nonumber \\
&& \times\left\{(1-z)^2\ac +\frac{s^2}{\mu^4}(1+z)^2\ap\right\}
\eeq
with $z=\z$, the index $B$ standing for the Born cross section,
$\alpha$ and $\beta$ are the electric and magnetic pion polarizabilities,
respectively, and $\mu$ is the pion mass. The second term in this
expression is equal to an interference of the Born amplitude with
the pion polarizabilities.

In order to go out beyond the low-energy model let us construct
DRs for the amplitudes of the elastic $\g\pi$
scattering. We will consider the helicity
amplitudes $\mpp$ and $\mm$ which expressed through Prange's
amplitudes \cite{pran} $T_1$ and $T_2$ as
\be 
\mpp=-\frac{1}{2t}\left(T_1+T_2\right), \qquad
\mm=-\frac{T_1-T_2}{2[(s-\m)^2+st]}.
\ee
These amplitudes have no kinematical singularities and zeros
and define the cross section of the elastic $\g\pi$ scattering
as the following:
\be
\sig=\frac1{256\pi^2}\frac{(s-\m)^4}{s^3}\left\{(1-z)^2 |\mpp|^2
 +s^2(1+z)^2 |\mm|^2\right\}.
\ee

We construct for the amplitude $\mpp$ a DR
at fixed $t$ with one subtraction
\beq
\lefteqn{Re \mpp (s,t)=Re \overline{M}_{++}(s=\m,t)+B_{++}} \nonumber \\
 &&+\frac{(s-\m)}{\pi}P~\int\limits_{4\m}^{\infty}d\sp~Im\mpp(\sp,t)\left[
\frac{1}{(\sp-s)(\sp-\m)}-\frac{1}{(\sp-u)(\sp-\m+t)}\right]
\eeq
where $B_{++}$ is the Born term equal to
\be
B_{++}=\frac{2e^2\m}{(s-\m)(u-\m)},
\ee
and $Re\overline{M}$ is expressed through the difference of the pion
polarizabilities as
\be
Re\overline{M}_{++}(s=\m,t=0)=2\pi\mu\ac .
\ee

Via the cross symmetry this DR is identical to a DR with two subtraction.
The subtraction function $Re \overline{M}_{++}(s=\m,t)$
is determined with help of the DR at fixed $s=\m$ with one
subtraction where the subtraction constant is expressed because of (7)
through the difference $\ac$:
\beq
\lefteqn{Re \overline{M}_{++}(s=\m,t) 
=Re\mpp(s=\m,t)-B_{++}(s=\m,t)=2\pi\mu\ac} \nonumber \\
&& +\frac{t}{\pi}\left\{P\int\limits_{4\m}^{\infty}
\frac{Im\mpp(\tp,s=\m)~d\tp}{\tp(\tp-t)}
- P\int\limits_{4\m}^{\infty}
\frac{Im\mpp(\sp,u=\m)~d\sp}{(\sp-\m)(\sp-\m+t)}\right\}.
\eeq

The DRs for the amplitude $\mm(s,t)$ have the same expressions (5) and (8)
with substitutions: $Im\mpp \to Im\mm$, $B_{++} \to B_{+-}=B_{++}/\m$ and
$2\pi\mu\ac \to 2\pi/\mu \ap$.

The dispersion sum rule (DSR) for the polarizability difference $\a$
can be obtained from the DR at fixed $u=\m$ without
subtraction for the amplitude $\mpp$ \cite{rad}:
\be
\a=\frac{1}{2\pi^2\mu}\left\{\int\limits_{4\m}^{\infty}~\frac{
Im\mpp(\tp,u=\m)~d\tp}{\tp} +\int\limits_{4\m}^{\infty}~\frac{
Im\mpp(\sp,u=\m)~d\sp}{\sp-\m}\right\}.
\ee

The DSR for the sum of the polarizabilities is
\be
\b=\frac{\mu}{\pi^2}\int\limits_{4\m}^{\infty}~\frac{Im\mm(\sp,t=0)~d\sp}
{\sp-\m}=\frac{1}{2\pi^2}\int\limits_{\frac32 \mu}^{\infty}~
\frac{\s_T(\nu)~d\nu}{\nu^2}
\ee
where $\s_T$ is the total cross section of $\g\pi$ interaction, $\nu$
is the photon energy in lab. system.

The DRs and DSRs for the $\gp$ scattering are saturated by
the contributions of the $\rho(770)$, $b_1(1235)$, $a_1(1260)$
and $a_2(1320)$ mesons
in the $s$ and $u$ channels and $\s$, $f_0(980)$ and $f_2(1270)$ mesons
in the $t$ channel. In the case of the process $\gg$ the DRs and DSRs are
saturated by the contributions of the $\rho(770)$, $\omega(782)$ and
$\phi(1020)$ mesons in the $s$ and $u$ channels and the $\s$, $f_0(980)$ 
and $f_2(1270)$ mesons in the $t$ channel.

\section{Analysis of the process $\gg$}
\label{sec:3}

The parameters of the $\rho$, $\omega$, $\phi$, $b_1$ and $a_2$ mesons are
given by the Review of Particle Properties \mbox{\cite{part}}.
The parameters of the $f_0$ and $f_2$ mesons are taken from the work
\cite{mars}. For the $a_1$ meson we take: $m_{a_1}=1230$ MeV,
$\G_{a_1}=450$ MeV and $\G_{a_1\to\g\pi}=640$ keV.
The parameters of the $S$-wave $\pi\pi$ interaction are not known.
The main contribution to this interaction could be
given by the $\s$ meson.

The light scalar isoscalar meson ($\s$ meson) with a mass
approximately equal to twice the constituent mass of u and d quarks
and with a big width is predicted to exist by a number of models for
the dynamical breakdown of chiral symmetry. Such a particle is seen
to occur, for example, within QCD adaptation of the Nambu-Jona-Lasinio
model \cite{namb}.
In the work \cite{shur} one has shown that the
instanton vacuum structure predicted for QCD necessarily contains also
such a meson in the 500--600 MeV range. In the last years  search for
the $\s$ meson was carried out in a number of works
\cite{svec,torn,ishida,aleks}. However, experimental evidence for
the $\s$ meson is still both equivocal and controversial.

In the present work we will consider the $\s$ meson as an effective
description of the strong $S$-wave $\pi\pi$ interaction using the
broad Breit-Wigner resonance expression.
The parameters of such a $\s$ meson are found from a fit to
the experimental data \cite{mars} for the $\gg$ process in the energy range
of $\sqrt{t}=270\div 825$ MeV (where $t$ is the square of total
energy in $\g\g$ c.m.s.).
For this reaction the Born term is equal to zero and the main contribution
is determined by the $S$-wave of $\pi\pi$ interaction. So, this process 
gives a good possibility to investigate such a contribution.
As the elastic $\g\pi$ scattering and the process $\g\g\to\pi\pi$
should be described by a common analytical function, we will use the same
DRs for the description of these both processes.

There are five free parameters: the mass, the full width and the decay
width into $\g\g$ of the $\s$ meson and the sum and the difference of the
$\pi^0$ meson polarizabilities. Therefore, we consider the following
variants of fitting:

a) For the first step we do a fit to the data \cite{mars} in the energy
region from 270 MeV up to 2 GeV using all five parameters. This fit gives
$\as=0.98\pm0.03$ and \mbox{$\ab=$}$-1.6\pm2.2$ (in units 
of $10^{-4}$fm$^3$). Then we repeat the fit in the energy range of
$\sqrt{t}=270\div 820$ MeV using the found values for the sum and the
difference of the polarizabilities.

b) The sum and the difference of the polarizabilities are fixed according to
two loops prediction of the ChPT \cite{bell}:
$\as=1.15\pm 0.30$, $\ab=-1.90\pm 0.20$.

c) The values of these sum and difference are taken from the works 
\cite{kal,kp}: $\as=1.00\pm 0.05$, $\ab=-0.6\pm 1.8$.

Results of fitting in the energy region from $\sqrt{s}=270$ MeV up to
825 MeV are presented in Fig.1 by the solid line (the variant "a"),
the dashed line (the variant "b") and the dotted one (the variant "c").
The dashed-dotted line shows the result of the work \cite{donog} of Donoghue
and Holstein.
The dashed-double dotted line is the result of the two-loop calculations
in the framework of the ChPT \cite{bell}.
As it is evident from this figure, our model essentially improved
the agreement with the data \cite{mars}.

The found parameters of the $\s$ meson are listed in table 1.
The values of these parameters, in particular, the full width and the decay
width into $\g\g$, depend strongly on the value of the sum $\as$.
The obtained values of the $\s$ meson parameters indicate, probably, that in
this case we should consider the $\s$ meson as an effective account of the big
contribution of the non-resonant $S$-wave $\pi\pi$ interaction rather than as 
a real particle.

\begin{table}
\centering
\caption{
Parameters of the $\s$ meson obtained from different variants of the
fit to the experimental data \cite{mars} of the process $\gg$}
\begin{tabular}{|c|c|c|c|c|c|c|} \hline
  &$m_{\s}$(MeV)&$\G_{\s}$(MeV)&$\G_{\s\to\g\g}$(keV)&$\chi^2$&$\as$&
$\ab$ \\ \hline
a &$547\pm45$&$1204\pm362$&$0.62\pm0.19$& 0.30& $0.98\pm0.03$&$-1.6\pm2.2$
\\ \hline
b &$471\pm23$&$706\pm164$&$0.33\pm0.07$& 0.42 &$1.15\pm0.30$&$-1.9\pm0.2$
\\ \hline
c &$584\pm32$&$1378\pm277$&$0.83\pm0.16$&0.31&$1.00\pm0.05$&$-0.6\pm1.8$
\\ \hline
\end{tabular}
\end{table}

The value of the sum of the $\pi^0$
meson polarizabilities, found in the variant "a"
from the fit to the experimental data \cite{mars} on the total cross section
of the process $\gg$ in the energy region up to 2 GeV, practically
coincides with the result obtained by Kaloshin et al. \cite{kp} from the
analysis of angular distributions of the reaction under consideration
in the region of the $f_2(1270)$ resonance. A very high sensitivity of the
results of our calculations in this energy region to the value of $\as$
led to the small value of its error.

The obtained value of the difference $\ab$ agrees within the errors
with the result of the work \cite{kal} where a simultaneous fit to data
\cite{mars} and \cite{boy} was used.
The big value of the error in this case is caused by of a weak dependence
of our calculation results on the value of $\ab$.

The found values of $\as$ and $\ab$ are consistent within the error
bars with the two-loop prediction of the ChPT \cite{bell}.

Fig.2 demonstrates a description of the experimental data \cite{mars} of
the process under investigation in the energy region up to
$\sqrt{t}\approx 2$ GeV using the found values of the $\s$ meson parameters
and the corresponding values of the $\as$ and $\ab$. The solid, dashed and
dotted lines in this figure show results obtained for the "a", "b" and
"c" variants, respectively.
The result of calculations essentially depends on the value of $\as$.
The dashed-dotted line in Fig.2 corresponds to the values of the
polarizabilities predicted in the quark confinement model \cite{im}:
$\as=0.45$, $\ab=1.05$. This result is in complete disagreement with the
experimental data in the region of the $f_2(1270)$ resonance.

The calculation of $\ab$ with the help of the DSR (9) for the variant "a"
gives:

\begin{center}
\begin{tabular}{rcccccl}
          &$\rho$ &$\omega$&$\phi$  &$f_0$  &$\s$    &    \\
$\ab^{DSR}$=&-1.79&-11.69&-0.04&+0.44&+10.07&$=-3.01\pm2.06$ .\\
\end{tabular}
\end{center}

The indicated error for this value is caused by the
errors for the $\s$ meson parameters only. 
The integration in this DSR was performed up to 25 GeV$^2$.
The contribution into the error
from a finite limit of the integration is about 10\% in this
case.
This value of $\ab$ does not conflict within the errors with the one
used for the fit "a" and with the prediction of the ChPT \cite{bell}.

The application of this DSR for calculation of the difference of the
$\pi^{\pm}$ meson polarizabilities leads to the following value:

\begin{center}
\begin{tabular}{rrrrrrrl}
  &$\rho$&$a_1$&$b_1$&$a_2$&$f_0$&$\s$&  \\
$\ac^{DSR}=$&-1.2&+2.1&+0.9&-1.4&+0.4&+9.5&$=10.3\pm 1.9$ .\\
\end{tabular}
\end{center}
This result differs from the prediction of the ChPT \cite{kir,bur} 
($\sim 5.6$) and is close to the values obtained by DSRs \cite{rad,petr}
earlier. This difference might be caused by a slow convergence of the 
underintegral expressions in the DSR (9) and so an incorrect saturation of
this DSR. Therefore, a more detail investigation of
a saturation of this sum rule is necessary.

On the other hand, the DSR (10) has good enough convergence of the 
underintegral expression and gives the values of the sum polarizabilities 
\cite{rad,petr} close to experimental ones.

It is worth noting that contrary to the DSR (9) the underintegral expressions
in the DRs (5) and (8) converge, via of the subtractions,
essentially more quickly and there is no
such a problem with a saturation of these DRs. As a result, such DRs allow
to obtain a good description of the experimental data \cite{mars} on the $\gg$
process in the energy region up to 2 GeV (Fig.2).

\section{Analysis of Compton scattering on the $\pi^{\pm}$ meson}
\label{sec:4}

The calculation of the $\gp$ back scattering cross section with the help of 
the DRs (5)
and (8) gave practically identical result for the "a", "b"  and "c" variants 
of the fit to the data \cite{mars} for the process $\gg$ (see Sec.~\ref{sec:3}).
The maximum difference between these variants is realized at $\sqrt{s}$=1GeV
and is smaller than 5\%. The results of the calculation for the variant "a" 
at $\ap$=0.22 and $\ac$=5.6 are shown in Fig.3 by the solid 
line. The dashed line corresponds to $\ac=0$. The dotted one is the 
contribution of the Born term + polarizabilities. The latter is identical 
to the low energy expression (2).
It is evident from this figure that the  correction to the low energy
expression for the back scattering is important in a wide energy region.
The analysis of the contribution of separate resonances showed, that
the contribution of the $\rho$ meson is essential only at the energy
$\sqrt{s}>500$ MeV, while the contribution of the $\s$ meson is big
in almost all considered energy region. The contribution of the $b_1$
and $a_2$ mesons is very small and even in the energy region of
850--1000 MeV it does not exceed 3.5\%.
The contribution of the $a_1$ meson becomes important at the energy higher 
than 750 MeV. This contribution increases from 8\% up to 50\% in the energy 
interval 750 -- 1000 MeV. Therefore, the data for the back Compton scattering
on the charged pion at such energies can be used, in particular, to determine
the decay widths of the $a_1$ meson.

For the $\gp$ forward scattering the $\s$ meson does not give the
contribution.
In this case the cross section is determined by the contribution of
the Born term, the sum $\ap$ and the $\rho$, $a_1$, $b_1$ and $a_2$ mesons.
The results of calculation of the forward scattering cross section
are given in Fig.4 where the solid line corresponds to account of
all contributions with $\ap$=0.22, the dashed and dashed-dotted lines show
the results of the similar calculation
but with $\ap=0$ and with the contribution of the $a_1$, $b_1$ and $a_2$ 
mesons equal to zero, respectively. The dotted line is contribution of the
Born+$\ap$. This figure demonstrates the big contribution of the sum
$\ap$ in the energy region near 1GeV which increases with the energy
in both relative and absolute values. 
Corrections to the low energy expression
(2) are smaller than 1\% at $\sqrt{s}<500$ MeV and become essential
at higher energies. The main contribution into these corrections is
given by the $\rho$ meson. The contributions of the $a_1$, $b_1$ and
$a_2$ mesons are small at $\sqrt{s}<850$ MeV. They grow from 2--4\%
at $\sqrt{s}=850$ MeV up to 12--16\% at $\sqrt{s}=1000$ MeV.

The Fig.5 shows the result of calculation of the total cross section
for the Compton scattering on the $\pi^{\pm}$ meson (solid line).
The dashed line is the cross section at $\alpha=\beta=0$ and dotted
one is the contribution of the Born+polarizabilities. This figure
demonstrates the possibility of the use of the data on the total
cross section of the elastic $\g\pi^{\pm}$ scattering at the energy
region than higher $\sim 600$ MeV for the extraction of the pion 
polarizability.
In the energy region up to 700 MeV the total cross section is dominated by
the low energy approach. In particular, at $\sqrt{s}=350$ MeV the correction
to this approach is equal to about 0.6\% what is in agreement with
the two-loop ChPT calculation \cite{bur}.

The relative contributions of $\ac$=5.6 and $\ap$=0.22 into the back and
forward cross sections, respectively, and their contribution into the total
cross section as a function of the energy $\sqrt{s}$ are shown in Fig.6.
As follows from this figure the relative contributions of
the $\ac$ and $\ap$  into the back and forward cross sections grow with
the energy and in the region of 1GeV exceed 100\% and 300\%, respectively.
The contribution of the
polarizability into the total cross section is determined mainly by the
sum $\ap$ which also reaches 100\% in this energy region.
These results permit to determine the pion polarizabilities
with good enough accuracy from the experimental data for the elastic
$\gp$ scattering in the energy region under consideration.
As follows from above-made analysis
the most model independent result can be obtained in the energy region
up to 750--850 MeV.

To obtain an expression for the extraction of
 the reliable value of $\ac$ from the experimental data
for the $\gp$ back scattering
let us present the differential cross section of this process
as the following:
\be
\sig=\ac^2 A_1+\ac A_2+A_3
\ee
where
\beq
A_1&=&\frac1{64}\frac{(s-\m)^4}{s^3}\m(1-z)^2 ,\nonumber \\
A_2&=&\frac1{64\pi}\frac{(s-\m)^4}{s^3}\mu (1-z)^2\tilde M_{++},
\nonumber\\
A_3&=&\frac1{256\pi^2}\frac{(s-\m)^4}{s^3}\left\{(1-z)^2\left[
\tilde M_{++}^2+Im\mpp^2\right]+s^2(1+z)^2|\mm|^2\right\},
\eeq
$$
\tilde M_{++}=Re\mpp-2\pi\mu\ac.
$$
The amplitudes $\tilde M_{++}$ and $\mm$ can be calculated using the
upper constructed DRs.

As the first term in (10) is very small in a wide enough range of
the energy, it can be neglected and we have
\be
\ac=\frac{\displaystyle \frac{d\s_{\g\pi}^{exp}}{d\Omega}-A_3}{A_2}.
\ee
Here $d\s_{\g\pi}^{exp}/d\Omega$ is the experimental value of the
$\gp$ scattering
cross section. However, in the region of 1GeV the first term contribution
in (11) could be about 10\%. Therefore, to find the value of $\ac$ with
high enough accuracy we should use in this energy region the full equation
(11). In the case of the experiment \cite{ahr} the expressions (11) and (12)
must be integrated over $z$ from -1 up to -0.766.

For the forward scattering we write the cross section as
\be
\sig=\ap^2 B_1+\ap B_2+B_3
\ee
where
\beq
B_1&=&\frac1{64}\frac{(s-\m)^4}{s\m}(1+z)^2 ,\nonumber \\
B_2&=&\frac1{64\pi}\frac{(s-\m)^4}{s\mu} (1+z)^2\tilde M_{+-},
\nonumber\\
B_3&=&\frac1{256\pi^2}\frac{(s-\m)^4}{s^3}\left\{(1-z)^2|\mpp|^2
+s^2(1+z)^2\left[\tilde M_{+-}^2+Im \mm^2\right]\right\},
\eeq
$$
\tilde M_{+-}=Re\mm-\frac{2\pi}{\mu}\ap.
$$
Neglecting the first term in (13) we have
\be
\ap=\frac{\displaystyle \frac{d\s_{\g\pi}^{exp}}{d\Omega}-B_3}{B_2}.
\ee

\section{Conclusions}
\label{sec:5}

$\bullet$
The DRs at fixed $t$ with one subtraction have been constructed for the
invariant helicity amplitudes of the $\g\pi$ scattering. Using these
DRs for the description of the process $\gg$ we have obtained from the fit to
the experimental data \cite{mars} for this process
the sum  and difference of the $\pi^0$ meson polarizabilities:
($\as=0.98\pm0.03$, $\ab=-1.6\pm2.2$), and found
the parameters of the $S$-wave $\pi\pi$ interaction represented by
the broad Breit-Wigner resonance formula.

$\bullet$ The analysis of the calculation of the cross section of elastic
$\gp$ scattering showed that
the $S$-wave $\pi\pi$ interaction gives the significant
contribution into the cross section of the $\g \pi$ back scattering.
However, the value of this
contribution does not depend practically on the variant of fitting
of the process $\gg$ with the aim to find parameters of this interaction.

$\bullet$ 
One suggested the method of the extraction of the pion polarizabilities
from the Compton scattering on the pion data in a wide energy region 
which took into account corrections to the low energy expression.

$\bullet$ One found kinematical regions where the corrections to
the low energy expression either were very small or good
calculated. 

$\bullet$ The contributions of $\ac$ (into the back scattering) and $\ap$
(into the forward scattering and into the total cross section)
grow with the energy and exceed 100\% in the region of 1 GeV.
The most model independent result of the pion polarizabilities extraction
from the Compton scattering on the charged pion can be obtained in 
the energy region up to 750--850 MeV.

$\bullet$ The experimental data for the elastic $\gp$ scattering in the
energy region up to \mbox{$\sim 1$ GeV} together with the $\gg$ data could be
used both for a determination of the pion polarizability values 
and for study of the $S$-wave $\pi\pi$ interaction.

\vspace{0.5cm}
{\it Acknowledgments}. We gratefully acknowledge helpful discussions
with the members of the A2 Collaboration
at Mainz, in particular with J. Ahrens, R. Beck, D. Drechsel and
Th.~Walcher. We would also like to thank A. L'vov and V. Petrun'kin
for useful comments. We wish to express our gratitude for the
hospitality of the members of the Sonderforschungsbereich 201 at
Mainz.

\section*{Appendix}
The contributions of the vector and axial-vector mesons
($\rho$, $\omega$, $\phi$ and $a_1$) are
calculated with the help of the expression
$$
Im\mpp^{(V)}(s,t)=\mp s~Im\mm^{(V)}(s,t)
=\mp 4g_V^2 s\frac{\G_0}
{(m_V^2-s)^2+\G_0^2}  \eqno (A1)
$$
where $m_V$ is the meson mass, the sign "+" corresponds to the
 contribution of the $a_1$ meson and
$$
g_V^2=6\pi\sqrt{\frac{m_V^2}{s}}\left(\frac{m_V}{m_V^2-\m}\right)^3
\G_{V\to\g\pi},
$$
$$
 \G_0=\left(\frac{s-4\m}{m_V^2-4\m}\right)^{\frac32}\G_V m_V.
\eqno (A2)
$$
Here $\G_V$ and $\G_{V\to \g\pi}$ are the full width and the decay width into
$\g\pi$ of these mesons, respectively.

The $b_1$ and $a_2$ mesons give relatively small contribution in
the energy region under consideration. Therefore, we calculate
it using a narrow width approximation.
$$
Im\mpp^{(b_1)}(s,t)=s~Im\mm^{(b_1)}(s,t)=4g_b^2 \pi s\delta(s-m_b^2),
\eqno (A3)
$$

$$
Im\mpp^{(a_2)}(s,t)=-\frac12 g_a^2 \pi \left[ s^2-s(4\m-t)+\mu^4
-\frac{s(s-\m)^2}{2m_a^2}\right]\delta(s-m_a^2) ,  \eqno (A4)
$$

$$
Im\mm^{(a_2)}(s,t)=-\frac12 g_a^2 \pi \left[\m-t-\frac{(s-\m)^2}
{4m_a^2}\right]\delta(s-m_a^2)   \eqno (A5)
$$
where
$$
g_b^2=6\pi\left(\frac{m_b}{m_b^2-\m}\right)^3~\G_{b_1\to \g\pi^{\pm}},
\qquad
g_a^2=160\pi\left(\frac{m_a}{m_a^2-\m}\right)^5~\G_{a_2\to \g\pi^{\pm}}.
$$

For calculation of the contribution of the $\s$, $f_0$ and $f_2$ mesons
we use the following expressions:
$$
Im\mpp^{\s}(t,s)=\frac{g_{\s}\G_{0\s}}{(m_{\s}^2-t)^2+\G_{0\s}^2},
\qquad
Im\mpp^{f_0}(t,s)=\frac{g_{f_0}\G_{0f_0}}{(m_{f_0}^2-t)^2+\G_{0f_0}^2},
\eqno (A6)
$$

$$
Im\mm^{f_2}(t,s)=\frac{g_{f_2}\G_{0f_2}}{(m_{f_2}^2-t)^2+\G_{0f_2}^2}
\eqno (A7)
$$
where
$$
g_{\s}=8\pi\frac{m_{\s}+\sqrt{t}}{\sqrt{t}}\left(\frac{\frac23
\G_{\s\to\pi\pi} \G_{\s\to\g\g}}
{m_{\s}\sqrt{m_{\s}^2-4\m}}\right)^{\frac12},
\qquad
g_{f_0}=16\pi\left(\frac{\frac23 \G_{f_0\to \pi\pi} \G_{f_0\to \g\g}}
{m_{f_0}\sqrt{m_{f_0}^2-4\m}}\right)^{\frac12},
$$
$$
\G_{0\s}=\frac{\G_{\s}}2 (\sqrt{t}+m_{\s})\left(\frac{t-4\m}
{m_{\s}^2-4\m}\right)^{\frac12},
\qquad
\G_{0f_0}=\G_{f_0}m_{f_0} \left(\frac{t-4\m}
{m_{f_0}^2-4\m}\right)^{\frac12}, \eqno (A8)
$$

$$
g_{f_2}=160\pi\frac{m_{f_2}^{3/2}}{t(m_{f_2}^2-4\m)^{\frac54}}
\sqrt{\frac{D_2(m_{f_2}^2)}{D_2(t)}~\G_{f_2\to \pi\pi}
\G_{f_2\to \g\g}},
$$
$$
\G_{0f_2}=\G_{f_2}\frac{m_{f_2}}{\sqrt{t}}\left(\frac{t-4\m}
{m_{f_2}^2-4\m}\right)^\frac52 \frac{D_2(m_{f_2}^2)}{D_2(t)}.
\eqno (A9)
$$
The decay form factor $D_2$ is given according to Ref.\cite{mars}
$$
D_2(t)=9+3(qr)^2+(qr)^4, \qquad q^2=\frac14 (t-4\m) \eqno (A10)
$$
and the effective interaction radius $r$ is assumed to be 1fm.
The factor $(m_{\s}+\sqrt{t})$ at the relations for $g_{\s}$ and $\G_{0\s}$
is introduced to get a more correct expression for a broad Breit-Wigner
resonance.

\newpage
\begin{figure}
\epsfxsize=16cm
\epsfysize=18cm
\epsffile{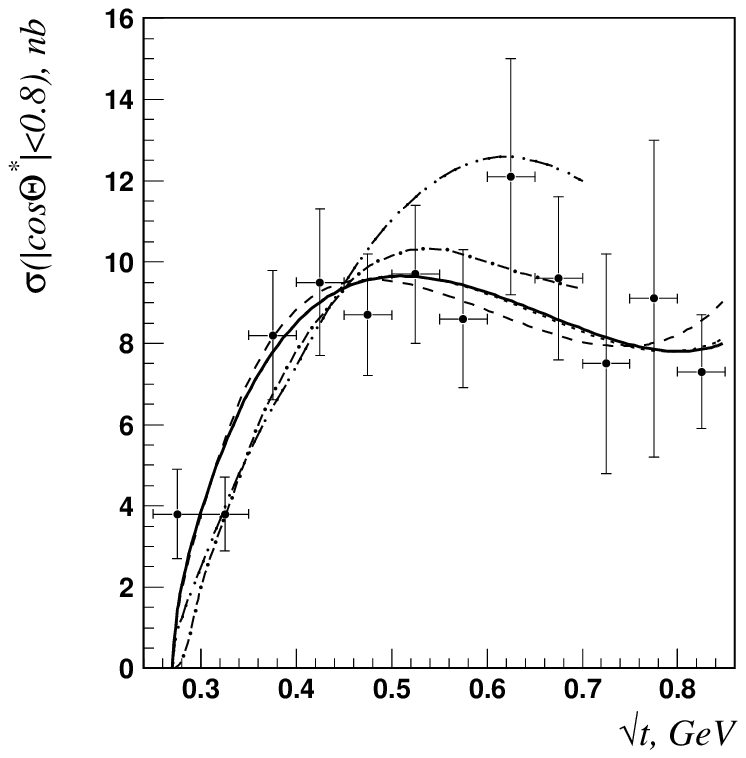}
\caption{
The cross section for $\gg$ process for $|\cos\theta^*|<0.8$,
where $\theta^*$ is the angle between the beam axis and one of the
$\pi^{0}$ in the $\gamma\gamma$ center-of-mass system.
The solid, dashed and dotted lines correspond to the "a", "b" and "c"
variants of fitting. The dashed-dotted and dashed-double doted lines are
the results of the work \cite{donog} and \cite{bell}, respectively.
Experimental data is taken from~\cite{mars}.}
\end{figure}

\newpage
\begin{figure}
\epsfxsize=16cm
\epsfysize=18cm
\epsffile{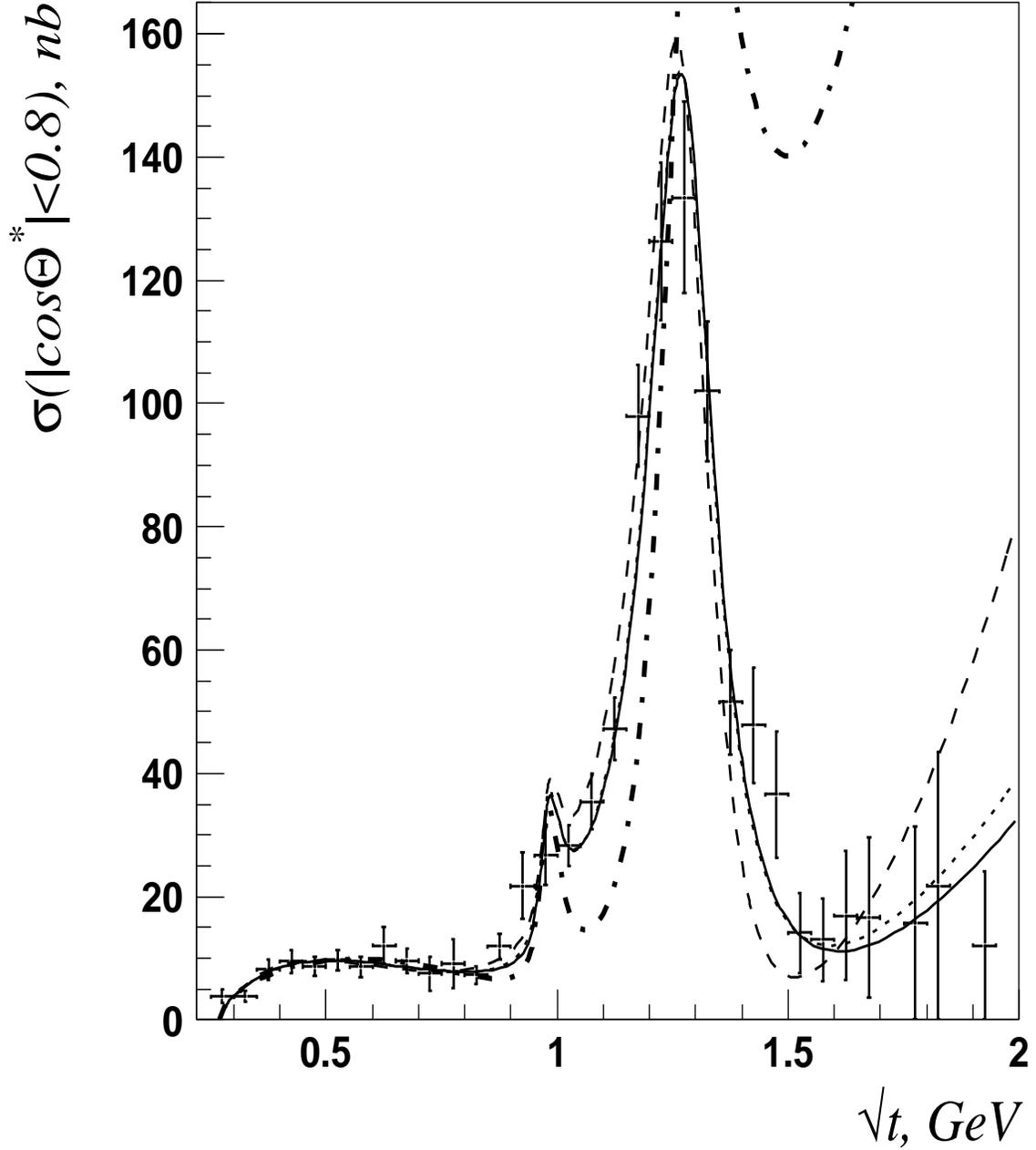}
\caption{
The cross section for $\gg$ process in the energy region up to 2 GeV.
The solid, dashed and dotted lines correspond to the "a", "b" and "c"
variants of fitting. The dashed-dotted line shows the result of calculations
with the values of the polarizabilities from the work \cite{im}.
Experimental data is taken from~\cite{mars}.}
\end{figure}

\newpage
\begin{figure}
\epsfxsize=16cm
\epsfysize=18cm
\epsffile{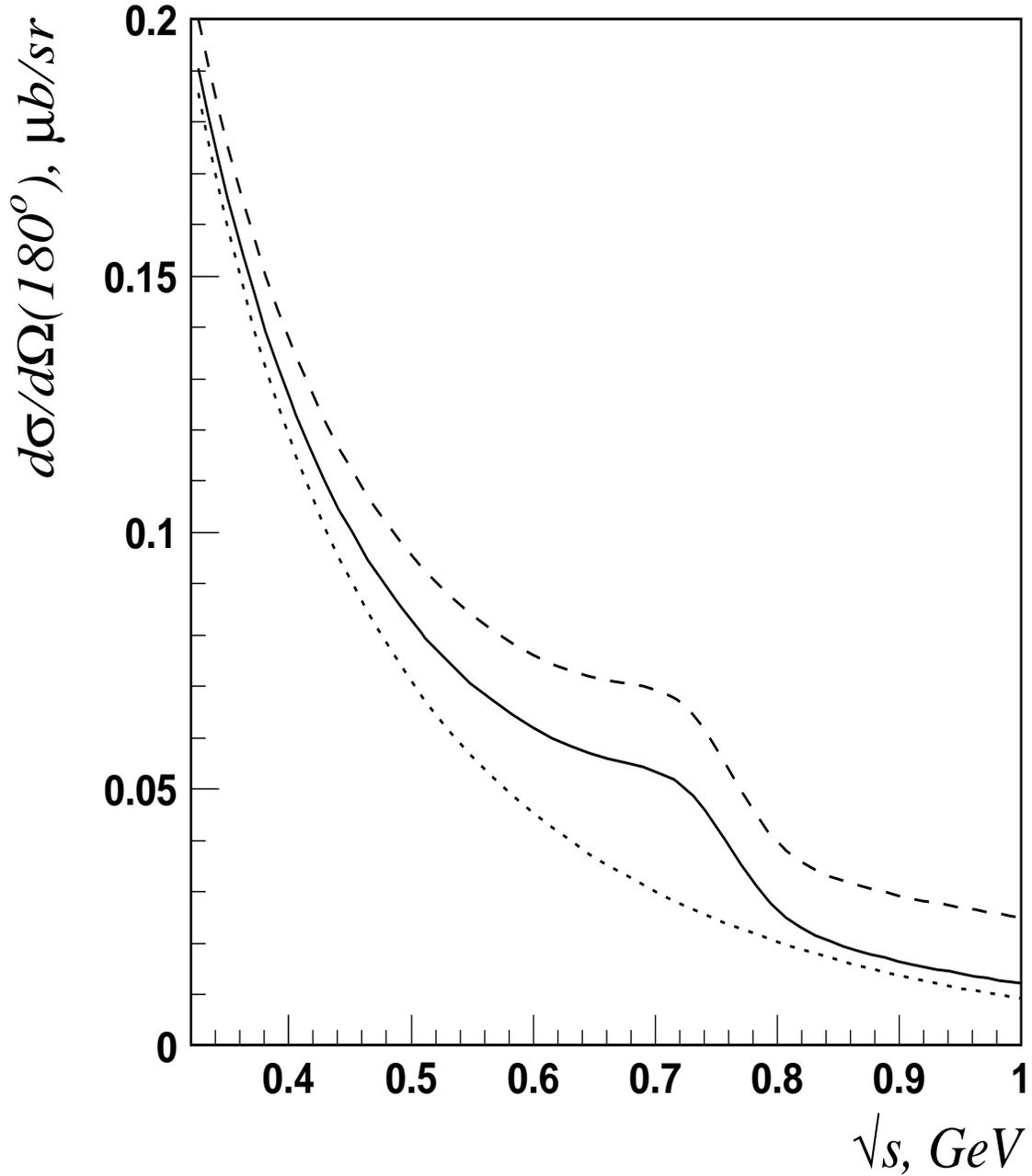}
\caption{
The back scattering cross section for $\gp\to\gp$ process.
The solid line is the result of calculation for the "a"
variant of the fitting at $\ac=5.6$, the dashed
line corresponds to $\ac =0$, the dotted one is the contribution of 
the Born term + $\pi^{\pm}$ meson polarizability.}
\end{figure}

\newpage
\begin{figure}
\epsfxsize=16cm
\epsfysize=18cm
\epsffile{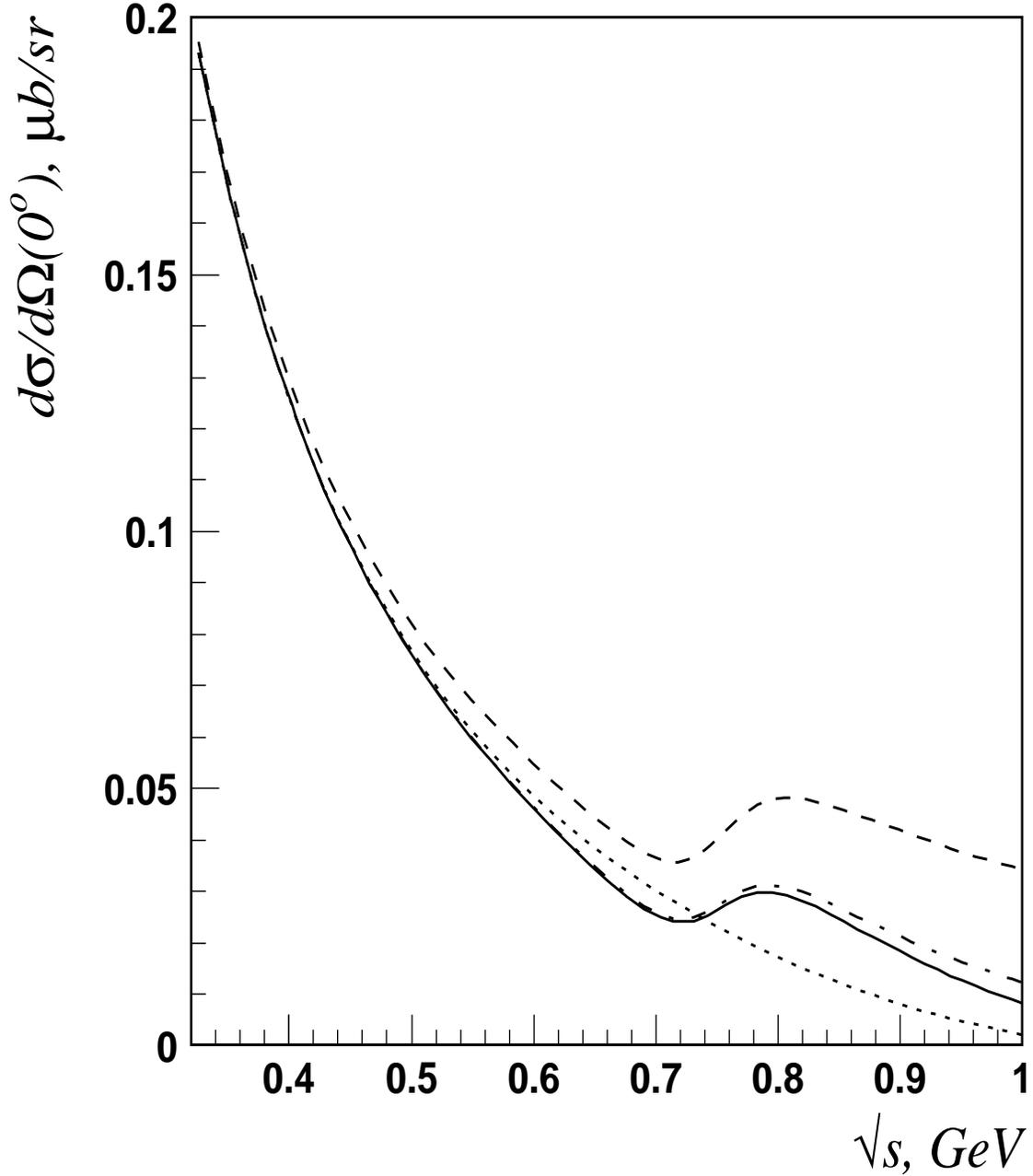}
\caption{
The forward scattering cross section for $\gp\to\gp$ process
at $\ap=0.22$ (solid line). The dashed line corresponds to $\ap =0$. The
dotted line is the Born term + polarizability contribution,
the dashed-dotted line corresponds to the contribution of the $a_1$, $b_1$ 
and $a_2$ mesons equal to zero.}
\end{figure}

\newpage
\begin{figure}
\epsfxsize=16cm
\epsfysize=18cm
\epsffile{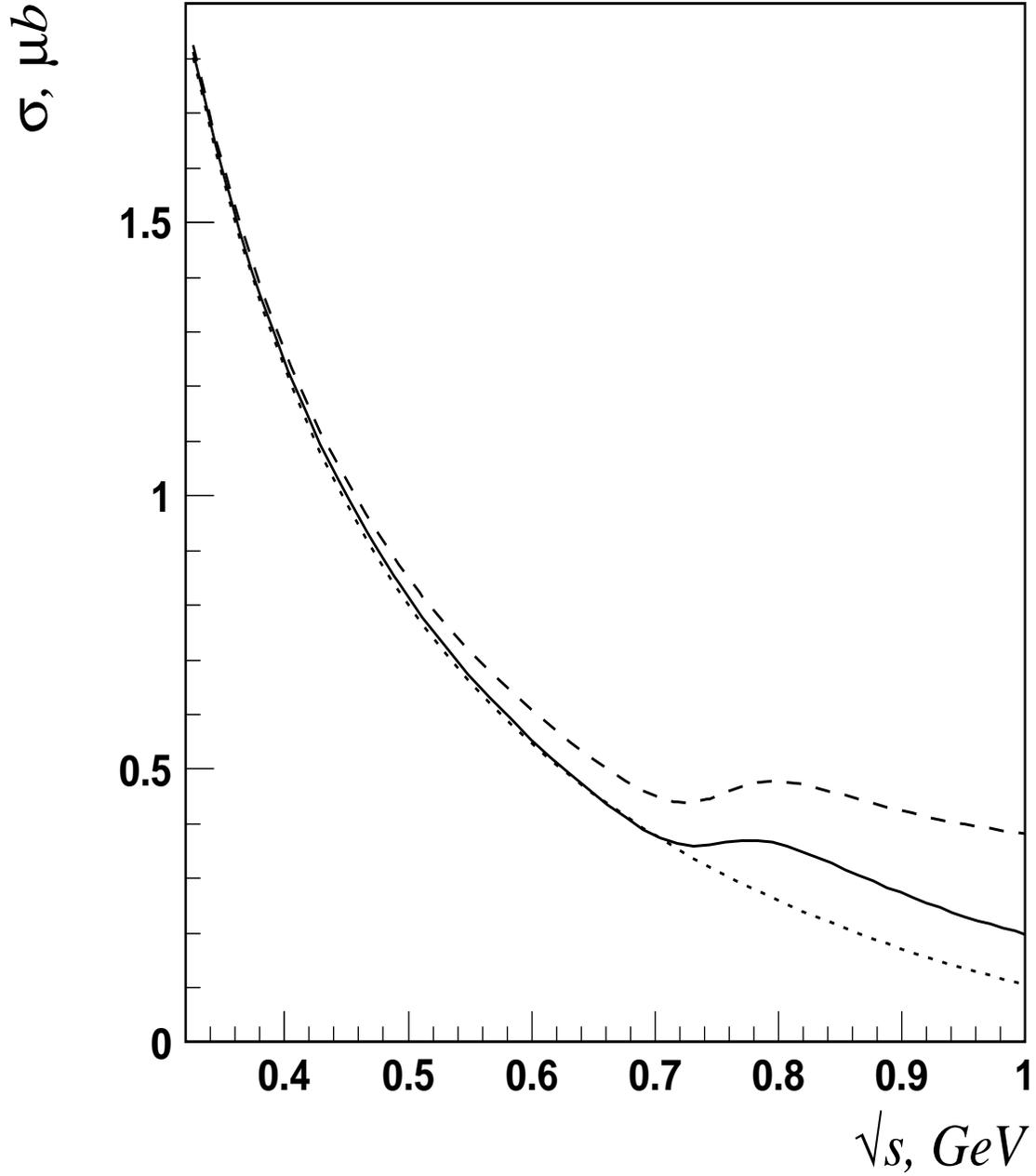}
\caption{
The total cross section for the elastic $\gp$ scattering
at $\ac=5.6$ and $\ap=0.22$ (solid line).
The dashed line is the cross section at $\alpha=\beta=0$ and dotted
one is the contribution of the Born+polarizabilities.}
\end{figure}

\newpage
\begin{figure}
\epsfxsize=16cm
\epsfysize=18cm
\epsffile{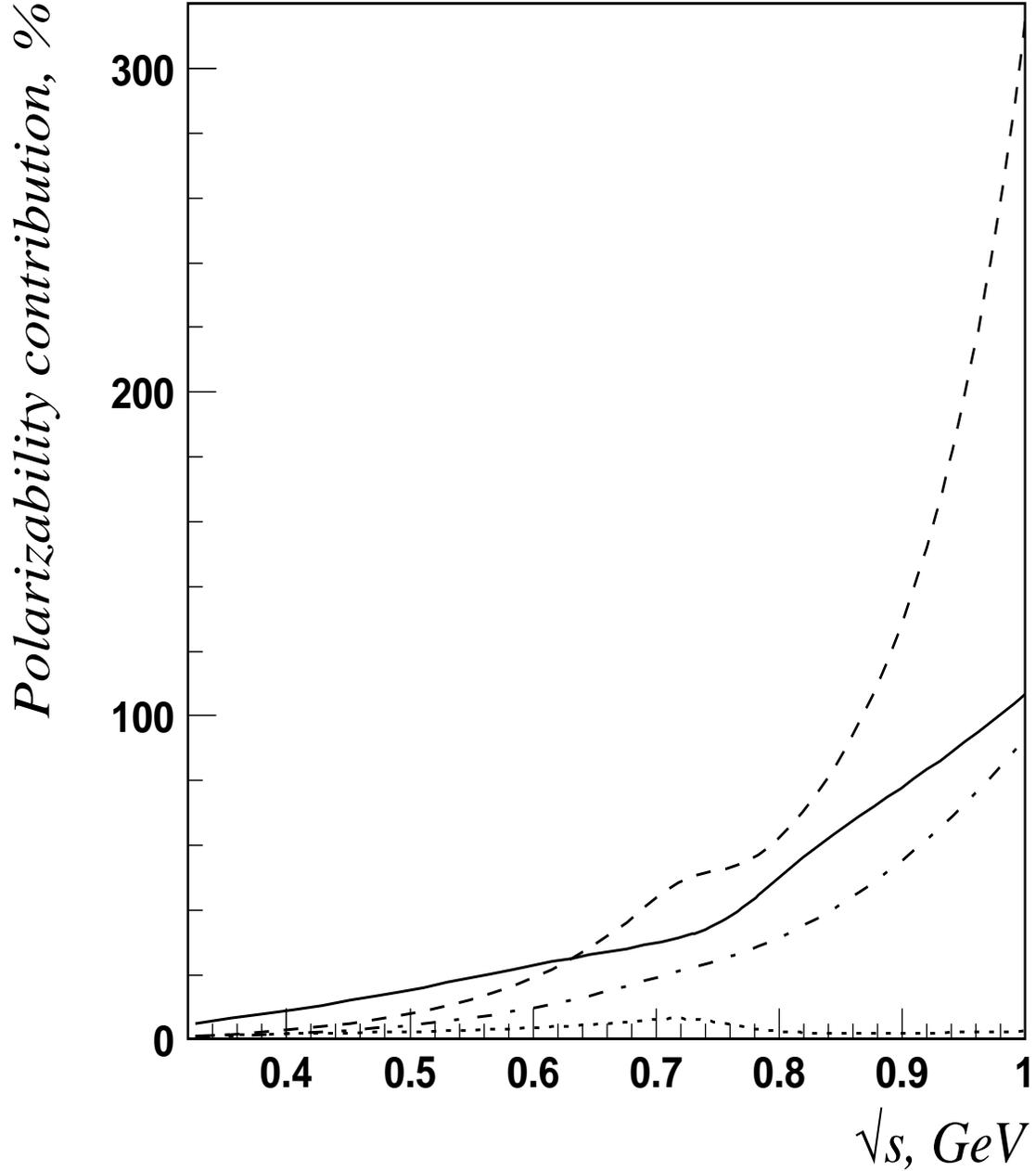}
\caption{
The relative contributions of $\ac$=5.6 (solid line)
and $\ap$=0.22 (dashed line) into the cross section for the back and forward
scattering, respectively. The dotted and dashed-dotted lines correspond
to the contribution of $\ac$ and $\ap$ into the total cross section of
the elastic $\gp$ scattering, respectively.}
\end{figure}

\end{document}